\documentstyle[twocolumn,epsf]{jpsj}
\def\runtitle{{\sc Letters}}
\def\runauthor{{\sc Letters}}
\def\lan{\left\langle}
\def\ran{\right\rangle}

\def\e{{\rm e}}
\def\virg{\;\;,}
\def\point{\;\,.}
\def\vf{v_{\rm F}}
\def\kf{k_{\rm F}}

\def\vrho{v_\rho}
\def\td{t_{\rm d}}
\def\ggs{\buildrel\textstyle > \over {\hbox{\raise0.2ex\hbox{$\sim$}}}}
\def\lls{\buildrel\textstyle < \over {\hbox{\raise0.2ex\hbox{$\sim$}}}}
\def\gsim{\,\lower0.75ex\hbox{$\ggs$}\,}
\def\lsim{\,\lower0.75ex\hbox{$\lls$}\,}

\title
{
Crossover from  Quarter-Filling to Half-Filling 
   in a One-Dimensional  Electron System
 with a Dimerized and Quarter-Filled Band%
 \footnote{submitted to J.~Phys.~Soc.~Jpn.}
 }

\author{
Masahisa {\sc Tsuchiizu},\hspace{-0.5mm}$^{1}$%
   \footnote{E-mail: tsuchiiz@edu2.phys.nagoya-u.ac.jp}%
   \hspace{0.5mm}
Hideo {\sc Yoshioka}$^2$ 
and 
Yoshikazu  {\sc Suzumura}$^{1,3}$ 
}

\inst{
$^1$Department of Physics, Nagoya University, Nagoya 464-8602 \\
$^2$Department of Physics, Nara Women's University, Nara 630-8506 \\
$^3$CREST, Japan Science of Technology Corporation  (JST)
}

\recdate{\hspace{3.5cm} }

\abst{
The interplay between 
quarter-filled  and half-filled umklapp scattering
   has been  examined  
   by applying the renormalization group method to  
 a one-dimensional quarter-filled electron system with 
   dimerization, 
 on-site ($U$) and nearest-neighbor ($V$) repulsive interactions. 
  The phase diagram   on the $U$-$V$ plane is obtained 
 at absolute zero temperature where 
  the Mott insulator (the charge ordered insulator) 
 is found for  smaller  (larger) $V$. 
By choosing the moderate  parameter in the  region 
   of Mott insulator,
it is shown  that the resistivity exhibits
   a crossover from 
 behavior of quarter-filling to that of  half-filling   
  with decreasing temperature.
}

\kword{quarter-filling, dimerization, 
       half-filling, umklapp scattering, Bechgaard salts
}

\begin{document}
\sloppy
\maketitle

The organic conductors, Bechgaard salts, 
   which show a crystal structure with a stacking 
   of organic molecules along one-dimensional axis, 
   form a quasi-one-dimensional system with a quarter-filled hole band
   \cite{Jerome,Gruner}.
Recently, the states above the phase transition temperature
   have been studied extensively to show  several experimental
   observations which may be  relevant to 
   the unconventional non-Fermi liquid.
Optical and photoemission experiments
   exhibit the charge gap in  both TMTTF salts 
   and TMTSF salts.
   \cite{Gruner_ICSM,Gruner_science}
The charge ordered (CO) insulating state 
   in (TMTTF)$_2$AsF$_6$ and (TMTTF)$_2$PF$_6$ 
   has been  maintained  
   by the NMR measurement\cite{Chow} 
   and  the huge dielectric response 
   in (TMTTF)$_2$PF$_6$ \cite{Monceau}
   indicates also the CO state.  
It is necessary to understand these experimental facts 
   in consistent with a  general phase diagram of the Bechgaard salts 
   indicating a dimensional crossover.\cite{Moser}

Two mechanisms for the insulating state has been  proposed
   theoretically \cite{Giamarchi_physica}.
One is of them is 
   the half-filled umklapp scatterings
induced by dimerization\cite{Emery} 
 and the charge gap has been examined  
 for the
  one-dimensional quarter-filled Hubbard model with 
  the dimerization.\cite{Penc}
The other is the commensurability at the quarter-filling.
  The quarter-filled extended Hubbard model
 without the dimerization  has been examined 
  by using a method of numerical diagonalization,
 where  the phase diagram shows  
   the insulating state with the charge gap  
   for a large  Coulomb repulsion
   \cite{Mila-Zotos,Sano-Ono}.
The similar phase diagram has been obtained analytically 
  by calculating  
   the magnitude of the 
   umklapp scattering at quarter-filling
   and  by using  the bosonization method and 
   the  renormalization group technique
   \cite{Yoshioka,Yoshioka2}.
 The model treating  both of these 
 two umklapp scattering on the same footing 
 is required  for  investigating   Bechgaard salts. 
  By  using the DMRG method, 
   the charge gap of  the model has been  estimated
   where  the metallic state   
   is reduced to the insulator in the presence of dimerization. 
   \cite{Nishimoto}.

In the present letter, 
  we investigate the interplay of the half-filled umklapp scattering
   and the quarter-filled one
   by  extending the previous work
 \cite{Yoshioka,Yoshioka2}
 to the case including  the dimerization. 
 It is demonstrated that 
   the competition between  these two scatterings gives rise to  
    two-kinds of unconventional insulating states in the ground state  
 and   that  a crossover  between these two states appears  
   in    the resistivity  at finite temperatures.

We consider 
   a one-dimensional quarter-filled extended Hubbard model 
   with dimerization.
The Hamiltonian is given by
\begin{eqnarray}
{\cal H} &=&
 - \sum_{j,\sigma} \left( t+ (-1)^j  \td \right)
   \left( c_{j,\sigma}^\dagger c_{j+1,\sigma} + {\rm h.c.}\right)
\nonumber \\
&&{}
 + U \sum_j n_{j,\uparrow} \, n_{j,\downarrow}
   + V \sum_j n_{j} n_{j+1}
\virg
\label{eq:H1D}
\end{eqnarray}
   where $n_{j,\sigma}=c_{j,\sigma}^\dagger c_{j,\sigma}$
   and $n_{j}= n_{j,\uparrow} + n_{j,\downarrow}$.
The fermion operator 
   $c_{j,\sigma}^\dagger$ denotes a creation
   of the electron at the $j$-th site with spin 
   $\sigma$. 
We have two kinds of transfer integrals $t+\td$ and $t-\td$ 
   due to  the dimerization.
The quantities $U$ and $V$ denote 
    the on-site and nearest-neighbor Coulomb repulsive interaction.
In  the presence of the dimerization, 
   the unit cell is given by  two lattice sites with  
    $2a$ where $a$ is the 
   lattice constant. 
 By  introducing operators  
   $c_{R_n, \sigma}^{A}$ and $c_{R_n+a,\sigma}^B$
   which are those  for the even and odd sites 
   in the $n$-th unit cell ($R_n=2an$), respectively, 
   we perform 
   the Fourier transform 
   $c_{k, \sigma}^A = N_0^{-1/2} \,
             \sum_{n}^{N_0} \, \e^{-i k R_n} \, c_{R_n,\sigma}^A$ 
   and 
   $c_{k, \sigma}^B = N_0^{-1/2} \,
             \sum_{n}^{N_0} \, \e^{-i k (R_n +a)} \,
    c_{R_n+a,\sigma}^B$, 
  where $N_0=L/(2a)$ and  $L$ is the length of the system.
Then the kinetic term of the Hamiltonian,
   ${\cal H}_0$, is diagonalized as
   ${\cal H}_0 =
   \sum_{k,\sigma} \varepsilon_k \,
          [  d_{k,\sigma}^\dagger d_{k,\sigma}
           - u_{k,\sigma}^\dagger u_{k,\sigma}]$
   with a dispersion relation, 
   $\varepsilon_k = -2[t^2\cos^2ka+\td^2\sin^2ka]^{1/2}$.
In a diagonalized basis,
   the lower  and upper band fermion operators are 
   given by 
   $ d_{k,\sigma} =
     [  \e^{ i\theta_k} \, c_{k,\sigma}^A 
      + \e^{-i\theta_k} \, c_{k,\sigma}^B ]/\sqrt{2}$ 
   and 
   $ u_{k,\sigma} =
     [ \e^{ i\theta_k} \, c_{k,\sigma}^A 
      - \e^{-i\theta_k} \, c_{k,\sigma}^B ]/\sqrt{2}$, respectively,
   where $\tan 2\theta_k = -(\td/t) \tan ka$.
Since a gap $ 4\td$ appears at $k=\pm \pi/2a$ and   
   the first Brillouin zone is reduced to half, 
   the system can be regarded effectively as half-filling 
   with  $\kf = \pi/4a$.

\begin{figure}[t]
\begin{center}
\leavevmode
\epsfxsize=7.5cm\epsffile{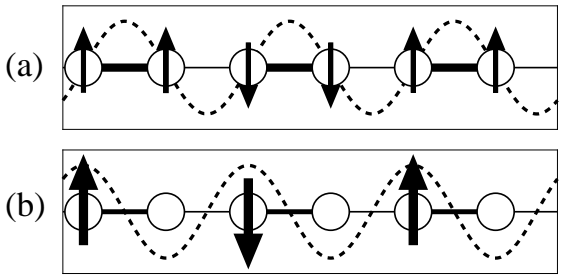}
\end{center}
\vspace*{-.5cm}
\caption{
Schematic view of the two limiting cases for 
    the insulating state:
   the Mott insulating state (a) and the CO insulating state (b).
The dashed wave indicates the $4\kf$ charge density wave with
   $\theta_\rho = \pi/4$ (a) and 
   $\theta_\rho = 0$ or $\pi/2$ (b).
}
\label{fig:config}
\end{figure}
In order to obtain the effective Hamiltonian for the states
   near $\pm \kf$, one  integrates out the 
   contribution from the upper band, 
which leads to the quarter-filled umklapp scattering\cite{Yoshioka}. 
We apply the bosonization method to 
   electrons in the lower band
   with  the dispersion  linearized as 
   $\varepsilon_k \to \pm \vf (|k|-\kf)$
 where  $\vf = \sqrt{2}ta [1-(\td/t)^2]/[1+(\td/t)^2]^{1/2}$.
Then the Hamiltonian, which  is rewritten 
   in terms of the charge and spin phase variable,
   is decoupled into two parts of 
   the charge and spin degrees of freedom.  
The charge part, ${\cal H}_\rho$, is expressed as \cite{Tsuchiizu_PHD}
\begin{eqnarray}
{\cal H}_\rho &=& 
\frac{v_\rho}{4\pi} \int {\rm d}x 
  \left[\frac{1}{K_\rho} \left(\partial_x \theta_\rho \right)^2
         + K_\rho \left(\partial_x \phi_\rho \right)^2
  \right]
\nonumber \\
&& {}
- \frac{g_{1/2}}{2 \pi^2 \alpha^2} \int {\rm d}x \, \sin 2 \theta_\rho
\nonumber \\
&& {}
+ \frac{g_{1/4}}{2 \pi^2 \alpha^2} \int {\rm d}x \, \cos 4 \theta_\rho
\virg
\label{eq:phaseHamiltonian}
\end{eqnarray}
   where $\alpha$ is a cutoff 
   of the order of lattice constant and 
    $[\theta_\rho(x), \phi_\rho(x')]= i \pi \, {\rm sgn}(x-x')$. 
In eq.~(\ref{eq:phaseHamiltonian}),
   the charge velocity and the Tomonaga-Luttinger exponent are
   given by
   $v_\rho = \vf 
    [(1+g_{4\rho}/2\pi\vf)^2-(g_\rho/2\pi\vf)^2]^{1/2}$ and
   $K_\rho = 
    [(2\pi\vf+g_{4\rho}-g_\rho)/(2\pi\vf+g_{4\rho}+g_\rho)]^{1/2}$.
The magnitude of half-filled umklapp scattering, $g_{1/2}$,
   and that of quarter-filled one, $g_{1/4}$, 
   are given by
\begin{eqnarray}
g_{1/2} &=&
 B 
\left[
Ua 
-  \frac{Aa^2}{2\pi\vf}  U   \left(U -2V \right) I_2(A)
\right]
\virg
\label{eq:g1/2}
\\
g_{1/4} &=& \frac{1}{(2\pi\alpha)^2} \, 
\frac{A^4 \, a^5}{\vf^2} \, U^2 (U-4V)
\virg
\label{eq:g1/4}
\end{eqnarray}
 where
   $A \equiv [1-(\td/t)^2]/[1+(\td/t)^2]$ and
   $B \equiv [2\td/t]/[1+(\td/t)^2]$.
  The quantity $g_{1/4}$  
  has been  derived 
  for  $\td=0$. \cite{Yoshioka} 
 The constants $g_\rho$ and $g_{4\rho}$
    are given by 
    \cite{Yoshioka}
\begin{eqnarray*}
g_{\rho} 
&=& (U + 4V)a 
\\ && {}
- \frac{Aa^2}{4\pi\vf}
  \Bigl[
     (U-2V)^2 
      + U^2 + (2V)^2
  \Bigr] \, I_1(A)
\virg
\\
g_{4\rho} 
&=& (U+ 4V)a 
\\  && {}
- \frac{Aa^2}{4\pi\vf}
  \Bigl[
     (U-2V)^2 
      + U^2  B^2
  + (2V)^2
  \Bigr]  I_2(A) 
\virg
\end{eqnarray*} 
 where 
   $I_1(A) \equiv  \int_0^{\pi/2} {\rm d}\varphi \,
   \sum_{\epsilon=\pm}
   [ 2+2\sqrt{1+\epsilon A \cos \varphi} ]^{-1}$,
 and 
   $I_2(A)\equiv  2\int_0^{\pi/2} {\rm d}\varphi \,\,
    [2+\sqrt{1+A \cos \varphi}+\sqrt{1-A \cos \varphi}]^{-1}$.
These quantities are calculated as 
   $I_1(A)\simeq I_1(1) = \ln(\sqrt{2}+1)$ and
   $I_2(A)\simeq I_2(1) = 2(\sqrt{2}+1)$.
The order parameters of the $4\kf$ charge density wave (CDW) state
   is given by\cite{Yoshioka}
\begin{eqnarray}
{\cal O}_{4\kf\mbox{-}\rm CDW}  
&\propto&
\cos \left( 4 \kf x +  2\theta_\rho \right)
\point
\end{eqnarray}

From eq.~(\ref{eq:phaseHamiltonian}), one  finds  the following 
   facts.
In the case of $g_{1/4}=0$,
   $\theta_\rho$ is  locked to  $\theta_{\rho}=\pi/4$ for $g_{1/2} > 0$,
  (or $\theta_{\rho} = 3\pi/4$ for $g_{1/2} < 0$)   
   with $0 \le \theta_{\rho} < \pi$.
In this  case, the charge configuration for $4\kf$-component is
   shown  schematically in Fig.~\ref{fig:config}(a),
   which corresponds  to the Mott insulating state.
 We note that the sign of the half-filled umklapp scattering
   depends on that of  the dimerization,
   i.e., $g_{1/2}>0$ for $\td>0$ and $g_{1/2}<0$ for $\td<0$.
On the other hand,
   in the case of $g_{1/2}=0$,
 the phase is locked to  
   $\theta_\rho = 0$, $\pi/2$
   for $g_{1/4} < 0 $ or 
  $\theta_\rho = \pi/4$, $3\pi/4$
   for $g_{1/4} > 0$.
The locking of $\theta_\rho$ for $g_{1/4}<0$ (i.e., 
   $U<4V$) corresponds to  the CO insulating state
   shown in Fig.~\ref{fig:config}(b). 
In the presence of both $g_{1/2}$ and $g_{1/4}$, 
 the behavior becomes  quite different 
depending on the sign of $g_{1/4}$.
In the case, $g_{1/4}>0$,
   the two kinds of umklapp scattering are compatible  and
   $g_{1/2}$-term removes the degeneracy 
   between $\theta_{\rho}=\pi/4$ and $\theta_{\rho}=3\pi/4$.
However, in $g_{1/4}<0$ case,
   the two kinds of umklapp scattering compete with each other 
   since the locking positions of $\theta_{\rho}$ are different 
   for respective umklapp scattering.
The CO state, which appears  in  large $V$,
   has been obtained by the mean-field approximation
   method \cite{Seo} 
   and the competition between these two kinds of potentials 
   has been   shown in terms of phase variable 
   \cite{Suzumura_JPSJ}.
Here we investigate    
   the  competition between   the Mott insulator 
   and the CO insulator, 
   by employing   the renormalization group method,  
   where the quantum fluctuation is fully taken into account.

The renormalization group equations 
 for the coupling constants  in  
 the Hamiltonian, eq.~(\ref{eq:phaseHamiltonian}), are given by 
\begin{eqnarray}
\frac{{\rm d}}{{\rm d}l} \, K_\rho 
&=& 
- 2 \, G_{1/2}^2 \, K_\rho^2
- 8 \, G_{1/4}^2 \, K_\rho^2
\virg
\\
\frac{{\rm d}}{{\rm d}l} \, G_{1/2}
&=&
\left(2-2K_\rho\right) G_{1/2} + G_{1/2} \, G_{1/4}
\virg
\label{eq:rg-g1/2}
\\
\frac{{\rm d}}{{\rm d}l} \, G_{1/4}
&=&
\left(2-8K_\rho\right) G_{1/4} + \frac{1}{2} \, G_{1/2}^2
\virg
\label{eq:rg-g1/4}
\end{eqnarray}
   where $G_{z}=g_{z}/(2\pi v_\rho)$ ($z =1/2$, 1/4) 
and $l=\ln(\alpha'/\alpha)$
   with the new length scale $\alpha'$ being larger than $\alpha$.
 We note that these renormalization group equations can be derived
   in a way similar  to Ref.~\citen{Tsuchiizu2}.
Equations~(\ref{eq:rg-g1/2}) and (\ref{eq:rg-g1/4}) 
   indicate the  fact that 
   $G_{1/2}$ and $G_{1/4}$ are suppressed by each other 
   for $G_{1/4}<0$ and are 
   enhanced  for $G_{1/4}>0$ as  discussed in the last paragraph.

\begin{figure}[t]
\begin{center}
\leavevmode
\epsfxsize=7.cm\epsffile{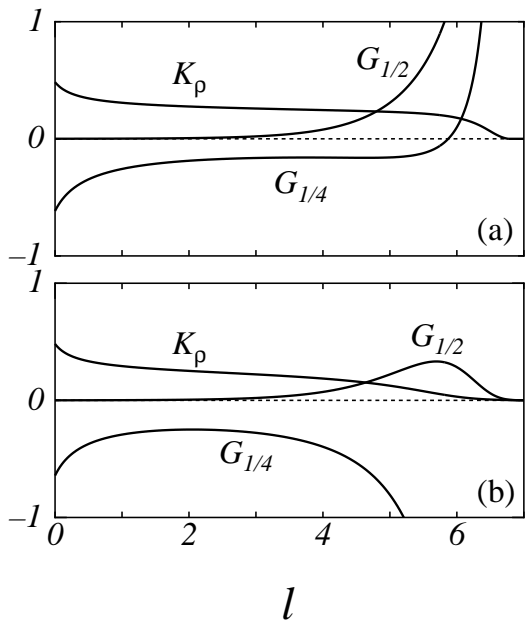}
\end{center}
\vspace*{-.5cm}
\caption{
The scaling flows of the coupling constants 
 for   $V/t=4.60$ (a) and 4.75 (b)
   where $U/t=6$ and $\td/t=0.001$. 
These are 
   the typical scaling flows for the Mott insulator (a) and 
   for the CO insulator (b), respectively.
}
\label{fig:RGflow}
\end{figure}
The competition for $g_{1/4} < 0$ 
 (i.e., $U < 4V$) is examined numerically. 
 The parameter of the dimerization is  taken as 
   $\td/t$ =0.001 to show results clearly, although   
 such a choice is rather small compared with that of 
  the organic conductors ($\td/t \simeq 0.1$).
In Fig.~1, the $l$-dependences of 
   $K_\rho(l)$, $G_{1/2}(l)$ and $G_{1/4}(l)$ are shown 
   for $U/t=6$ and $\td/t=0.001$ with fixed $V/t=4.60$ (a) and 4.75 (b).
For both cases, the exponent $K_\rho$
   decreases  monotonically  indicating
   the insulating state.
In Fig.~1(a), the half-filled umklapp scattering $G_{1/2}(l)$ 
   increases  and becomes relevant leading to the
   Mott insulator.
The sign of $G_{1/4}(l)$ changes from negative to positive 
 due to the relevant $G_{1/2}(l)$.
Figure~1(b) exhibits  $G_{1/2}(l)$ and $G_{1/4}(l)$, which  are 
   quite different compared with those  of Fig.~1(a). 
 In this case,    
   $G_{1/4}(l)$ is  relevant while  
   $G_{1/2}(l)$  
decreases after taking a maximum and becomes
   irrelevant.  
The relevance of $G_{1/4}$ with negative value denotes 
   the locking of $\theta_\rho$ being  $0$ or $\pi/2$
  and leading to   the CO insulating state.

\begin{figure}[t]
\begin{center}
\leavevmode
\epsfxsize=7.5cm\epsffile{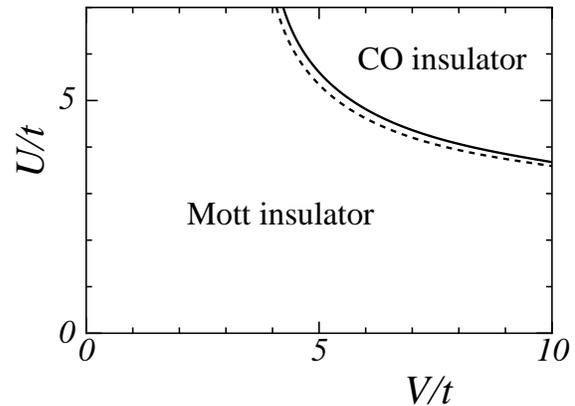}
\end{center}
\vspace*{-.5cm}
\caption{
The phase diagram of the Mott insulator and the CO insulator
   on the plane of $U/t$ and $V/t$ with fixed 
   $\td/t=0.001$.
The dashed curve indicates the boundary in the limit of $\td \to 0$. 
}
\label{fig:phase}
\end{figure}

Based on Fig.~2, 
 the phase diagram of the ground state
  is shown in Fig.~\ref{fig:phase} with fixed
   $\td/t=0.001$.
In the Mott insulator, 
   the phase $\theta_\rho$ is locked at  $\theta_\rho=\pi/4$, while
   in the CO insulator the phase is fixed at   
   $\theta_\rho=0$ or $\pi/2$.
The dashed curve denotes  the boundary in the limit of $\td\to0$,
   which is equal to that of the 
   metal-insulator transition in the quarter-filled extended 
   Hubbard model.\cite{Yoshioka}
In the presence of  $\td$, 
 the metallic region for $\td=0$
 changes  into the Mott insulating state and the CO insulating region 
   is suppressed.
 At the boundary, 
 the charge configuration 
 changes 
 from uniform (Fig.1(a)) to CO (Fig.1(b)) with increasing $V$.  
We note that the state at the boundary becomes  metallic 
 since this case corresponds to  an Ising transition 
 found  
    in the sine-Gordon system 
   with two kinds of non-linear terms, 
 i.e.,    the gap collapses on the criticality
   \cite{Delfino,Fabrizio,Orignac}.

Now we examine the competition at finite temperatures. 
   We calculate the temperature dependence of the 
   resistivity using the memory functional approach\cite{Mori,GW}
  which has been applied to the one-dimensional systems by 
  Giamarchi.\cite{Giamarchi_Resis}
The conductivity is given by
   $\sigma(\omega) = i [\chi(\omega)-\chi(0)]/\omega$,
   where $\chi$ is the retarded current-current correlation function:
   $\chi(\omega) \equiv \lan\lan j ; j \ran\ran_\omega \equiv
    -i \int dx \int_0^\infty dt 
    \lan \left[ j(x,t), \, j(0,0) \right] \ran $ 
   $e^{i\omega t-\delta t}$.
   The current operator $j(x,t)$ is given by
   $j(x,t) =  - \partial_t \theta_+(x,t)/\pi
   =-2  v_\rho K_\rho \, \Pi(x,t)$
   where $\Pi\equiv -\partial_x \theta_-/(2\pi)$.
 From the memory function approach,  
   $\sigma(\omega)$ is  rewritten as
   \cite{Giamarchi_Resis}
\begin{eqnarray}
\sigma(\omega) &=& 
i \, \frac{2\vrho K_\rho}{\pi} \, \frac{1}{\omega+M(\omega)} \virg
\end{eqnarray}
   where  $M(\omega)$ is  the memory function defined by 
   $M(\omega)\equiv \omega \, \chi(\omega)/[\chi(0)-\chi(\omega)] $.
 The memory function is calculated perturbatively as
   $M(\omega) \simeq -
   [\lan\lan F; F\ran\ran_{\omega}^0
                 -\lan\lan F; F\ran\ran_{\omega=0}^0 ]
    /[\omega\chi(0)]$
   where 
   $\lan\lan F; F\ran\ran_{\omega}^0$
   stands for the retarded correlation function 
  in the absence of umklapp scattering and $F\equiv [j,{\cal H}]$.
In terms of  $M(\omega)$,   
 the resistivity is expressed  as
\begin{eqnarray}
\rho(T) &=& \lim_{\omega\to 0} \sigma^{-1} (\omega)
= \frac{\pi}{i 2\vrho K_\rho} \lim_{\omega\to 0} M(\omega)
\point
\label{eq:resis}
\end{eqnarray}
Equation (\ref{eq:resis}) indicates the power-law behavior, 
\begin{eqnarray}
\rho(T) \propto \sum_{n=1,2} G_{1/2n}^2 \, T^{4n^2 K_\rho-3} 
\point
\end{eqnarray}
In order to obtain the precise behavior  of the resistivity,
   we  use  the solutions of  the renormalization group equations, 
   in which the umklapp scattering
   becomes relevant at low temperature. 
By putting $e^{-l}=2\pi T/W$ with $W\equiv \vrho \alpha^{-1}$, 
   the resistivity is obtained as 
\begin{eqnarray}
\rho(T) &=& 
\sum_{n=1,2} 
\frac{4n^2\pi}{\alpha} \,G_{1/2n}^2(l) \, 
  \cos^2\left(n^2\pi K_\rho(l)\right)  
\nonumber \\ && \hspace*{0.5cm} {} \times 
B^2\left(n^2 K_\rho(l),1-2n^2 K_\rho(l)\right) \, e^{-l}
\point
\end{eqnarray}
   where 
   $B(x,y)$ is the beta  function given by 
   $B(x,y)\equiv \Gamma(x)\Gamma(y)/\Gamma(x+y)$. 
\begin{figure}[t]
\begin{center}
\leavevmode
\epsfxsize=7.5cm\epsffile{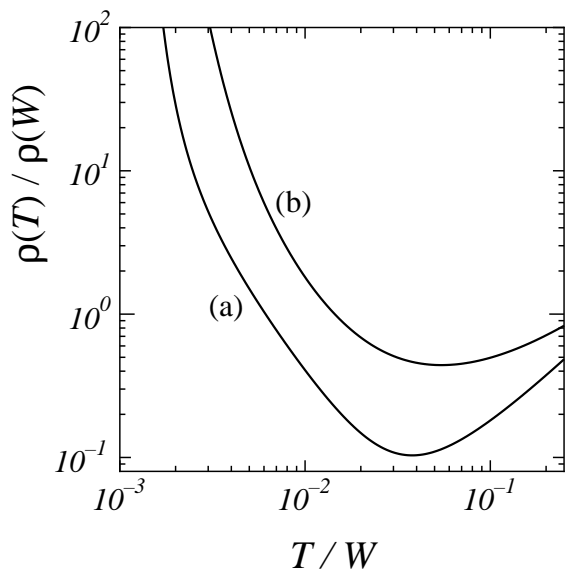}
\end{center}
\vspace*{-.5cm}
\caption{
The temperature dependence of the resistivity
  for $U/t=6$ and $\td/t=0.001$with fixed 
  $V/t=4.60$ (a) and 4.75 (b) where 
   $W$ denotes the cutoff 
   of the order of the bandwidth.
 The curve (a) (curve (b)) corresponds  
 to the behavior of the Mott insulator (the CO insulator).   
}
\label{fig:resistivity}
\end{figure}
In Fig.~\ref{fig:resistivity}, the temperature dependence of the
   resistivity is shown for $U/t=6$ and $\td/t=0.001$ with fixed 
   $V/t=4.60$ (a) and 4.75 (b).
The curve (a) shows  the behavior of the Mott insulator,
   while the curve (b) represents  that of the CO insulator.
For the CO insulator (curve (b)), 
    the resistivity is determined by  
   the quarter-filled umklapp scattering, where
   the resistivity exhibits the power-law dependence given by 
    $\rho(T)\propto T^{16K_\rho -3}$
    around the temperature $T/W\gsim 1\times 10^{-1}$.
  The temperature $T/W\simeq 5\times10^{-2}$, 
  corresponding to a minimum of   the resistivity,
   is of the order  of  the charge gap $\Delta_\rho$
   below which the resistivity increases rapidly.
   It is expected that the resistivity 
   for $T\ll\Delta_\rho$ behaves as $\exp(\Delta_\rho/T)$
   \cite{Giamarchi_physica}, 
 although  the perturbative renormalization group
   approach breaks down for $T\ll \Delta_\rho$. 
For the case of the Mott insulator (curve (a)),
we found a noticeable behavior due to  
  the relevance of the half-filled umklapp scattering. 
 There appears a mid  temperature region 
   around  $T/W\simeq 1\times 10^{-2}$, where
   the resistivity shows power-law behavior of half-filling 
 given by    $\rho(T)\propto T^{4K_\rho-3}$.
  At higher temperatures ($T/W\gsim 4\times10^{-2}$),
  the contribution of the quarter-filled umklapp scattering
  becomes dominant 
    at which the resistivity is similar to the case of curve (b).
 Such a crossover  in the region of CO insulator
 can be found by  the moderate choice of 
 parameter with  $|G_{1/4}| > G_{1/2}$, i.e., 
 close to the boundary in Fig.~3. 
The similar result is found in the frequency dependence of the 
   conductivity at $T=0$.
While  
the optical conductivity of  the CO insulating phase   
 corresponding to curve (b)
  shows 
   the power-law dependence with 
   $\sigma(\omega)\propto \omega^{16K_\rho-5}$ 
   above the frequency of  the charge gap, 
 the conductivity  of the Mott insulator corresponding to 
 curve (a) shows 
   a crossover  
   from $\sigma(\omega)\propto \omega^{16K_\rho-5}$
   to $\sigma(\omega)\propto \omega^{4K_\rho-5}$
   as decreasing frequency. 
Thus, the crossover from quarter-filling to half-filling 
 is found  
 with decreasing temperature or frequency 
 when parameters are located in the  Mott insulating region 
 and close to the boundary.

Finally we comment on  observation in the Bechgaard salts, 
   (TMTTF)$_2$X and (TMTSF)$_2$X.
 The optical measurement in the normal 
   states of these salts shows that 
   the optical conductivity has  a power-law dependence 
  as $\sigma(\omega)\propto\omega^{-1.3}$
   at  high frequency, 
  indicating a fact that   
the quarter-filled umklapp scattering
   is dominant
\cite{Gruner_Giamarchi}.
 Our calculation suggests that 
   the quarter-filled umklapp scattering becomes always  dominant
     at high frequency 
   when the parameter  is close to the boundary.
Therefore
   the power-law dependence of the optical conductivity 
   $\sigma(\omega)\propto \omega^{16K_\rho-5}$
  could be compatible with the charge gap, which  
  originates in   the Mott insulator and/or the CO insulator.  

In conclusion,
  we found   two kinds of insulator,
   the Mott insulator and the CO insulator,
   for  the ground state of 
 a  one-dimensional quarter-filled system with dimerization, 
  where 
    the half-filled umklapp scattering  induced 
   by the dimerization  competes with 
   the quarter-filled 
   umklapp scattering for large nearest-neighbor
   Coulomb interaction $V(>U/4)$.  
 In the Mott insulating phase close to the boundary, 
   the half-filled umklapp scattering is dominant at low $T$ but  
    the quarter-filled umklapp scattering 
   becomes  dominant at high $T$,
   and then the resistivity 
  could exhibit   a crossover from a quarter-filled behavior
   to a half-filled one as decreasing temperature.

\acknowledgements
The authors would like to thank 
   E.~Orignac for  discussion on the resistivity 
and also H.~Seo for useful comments.


\begin{thebibliography}{99}
\bibitem{Orignac}
The correspondence to the present model 
  has been pointed out by E. Orignac.
\end{thebibliography}
\end{document}